\documentclass[12pt]{iopart}
\usepackage{iopams, graphics, epsfig}

\def\beqa{\begin{eqnarray}}
\def\eeqa{\end{eqnarray}}
\def\beq{\begin{equation}}
\def\eeq{\end{equation}}

\def\l{\cal L}

\def\pa{\partial}

\def\alp{\alpha}

\def\bq{{\bf q}}
\def\bqd{{\bf \dot{q}}}
\def\qd{\dot{q}}

\def\eqdef{\buildrel {\rm def} \over =}
\def\({\left(}    \def\){\right)}

\def\data{\the\day-\the\month-\the\year}

\def\frac#1#2{{#1 \over #2}}

\def\phi{\varphi}

\def\~{\approx}
\def\pr{{\it Phys. Rev.}\ }

\def\pl{{\it Phys. Lett.}\ }

\def\cqg{{\it Class. Quantum Grav.}\ }

\def\grg{{\it Gen. Relativ. Grav.}\ }

\def\mnras{{\it Mon. Not. R. Ast. Soc.}\ }

\def\l{\cal L}

\begin{document}

\title{Spherically symmetric solutions in $f(R)$ gravity {\it via} Noether Symmetry Approach }

\author{S. Capozziello$^{1}$,
A. Stabile$^{2}$, A. Troisi$^{1}$}

\address{$^1$Dipartimento di Scienze Fisiche, Universit\`a di Napoli "Federico II",
and INFN, Sez. di Napoli,  Compl. Univ. di Monte S. Angelo,
Edificio G, Via Cinthia, I-80126 - Napoli, Italy}

\address{$^2$Dipartimento di Fisica ``E.R. Caianiello'' and INFN, Sez. di Napoli,
Universit\`a di Salerno, via S. Allende, I-84081 - Baronissi (Sa),
Italy}

\begin{abstract}
We search for spherically symmetric solutions of $f(R)$ theories
of gravity via the Noether Symmetry Approach. A general formalism
in the metric framework is developed considering a point-like
$f(R)$\,-\,Lagrangian where spherical symmetry is required.
Examples of exact solutions are given.

\end{abstract}


\section{Introduction}
Extended Theories of Gravity have become a sort of paradigm in
modern physics since they seem to solve several shortcomings of
standard General Relativity (GR) related to cosmology,
astrophysics and quantum field theory. The idea to extend
Einstein's theory of gravitation is fruitful and economic with
respect to several attempts which try to solve problems by adding
new and, most of times, unjustified new ingredients in order to
give a self-consistent picture of dynamics.  The today observed
accelerated expansion of Hubble flow and the missing matter of
astrophysical large scale structures, are primarily enclosed in
these considerations. Both the issues could be solved changing the
gravitational sector, i.e. the {\it l.h.s.} of field equations.
The philosophy is alternative to add new cosmic fluids (new
components in the {\it r.h.s.} of field equations) which should
give rise to clustered structures (dark matter) or to accelerated
dynamics (dark energy) thanks to exotic equations of state. In
particular, relaxing the hypothesis that gravitational Lagrangian
has to be a linear function of the Ricci curvature scalar $R$,
like in the Hilbert-Einstein formulation, one can take into
account, as a minimal extension, an effective action where the
gravitational Lagrangian is a generic $f(R)$ function. As further
request, one can ask for $f(R)$ being analytical in order to
recover, at least locally, the positive results of GR. Several
studies in this sense show that cosmic dynamics at early
\cite{starobinsky,kerner} and late \cite{noi} epochs can be
successfully reproduced. On the other hand, flat rotation curves
of spiral galaxies can be fitted adopting the low energy limit of
power law $f(R)$ and without considering huge amounts of dark
matter in galactic haloes \cite{mnras,jcap}. Interesting
indications have been achieved also for other $f(R)$ functions
\cite{navarro} and for large scale structure (this issue is still
matter of debate \cite{bean}, the CMBR\,-\,spectrum turns out to
be very slightly affected if the Lagrangian shows a small
deviation from the standard Hilbert-Einstein form \cite{barrow}).
Despite of these positive results, the problem is still open since
the degeneration of viable Lagrangians has not been removed yet
\cite{mimicking} and a final theory embracing the phenomenology at
local and large scales or at early and late epochs, considering
only the $f(R)$ approach, is not available up to now (see
\cite{noi-odin,amendola} for a recent discussion). On the other
hand, Solar System experiments are not giving univocal constraints
on the Parametrized Post-Newtonian limit of such theories: some
authors claim for the fact that there is room for Extended
Theories of Gravity considering experimental data \cite{arturo,
matteo,tegmark}, others claim for ruling out such theories with
respect to GR \cite{olmo,kamionkowski}. The debate essentially
lies on the physical meaning of the conformal transformations. For
a recent and illuminating discussion on the argument see
\cite{faraoni} and references therein. A part the controversies,
and the open discussions, several efforts have been done to give
self-consistent formulations of $f(R)$ gravity \cite{stelle} and
several approaches have been pursued to find out solutions of the
field equations coming out from such theories, both in metric and
in Palatini formalism.

In a recent paper \cite{multamaki}, spherically symmetric
solutions for $f(R)$ gravity in vacuum have been found
considering relations among functions defining the spherical
metric or imposing a constant Ricci curvature scalar. The authors
have been able to reconstruct, at the end, the form of some $f(R)$
theories, discussing their physical relevance. In
\cite{multamaki2}, the same authors have discussed static
spherically symmetric perfect fluid solutions for $f(R)$ gravity
in metric formalism. They showed that a given matter distribution
is not capable of determining the functional form of $f(R)$.

In this paper, we want to seek for a general method to find out
spherically symmetric solutions in $f(R)$ gravity and, eventually,
in generic extended theories of gravity. Asking for a certain
symmetry of the metric, we would like to investigate  if such a
symmetry holds for a generic theory of gravity. In particular for
the $f(R)$ theories. Specifically, we want to apply the Noether
Symmetry Approach \cite{noether-capoz} in order to search for
spherically symmetric solutions in generic $f(R)$ theories of
gravity. This means that we consider the spherical symmetry for
the metric as a Noether symmetry and search for $f(R)$ Lagrangians
compatible with it. The method can give several hints toward the
formulation of Birkhoff's theorem (see \cite{ellis} for a general
formulation) for these theories since, up to now, there are
controversial results in this direction\footnote{Some authors
state that the theorem is not valid in general \cite{clifton}
while others claims for its validity for specific classes of
$f(R)$ \cite{neville,yasskin,barraco}.}.

The layout of the paper is the following. In Sec.2, we derive and
discuss the field equations for $f(R)$ gravity. Sec.3 is devoted
to the construction of the point-like Lagrangian for a generic
$f(R)$ theory. We point out that imposing the spherical symmetry
in the action, and then deriving the Euler-Lagrange equations, is
equivalent to derive first the field equations and then to impose
the spherical symmetry. This  procedure\footnote{It is
straightforward to show that this method works also for
Friedmann-Robertson-Walker (FRW) metric or generic Bianchi's
metrics and it is extremely useful to find out cosmological
solutions \cite{deritis,noether-capoz,CQGRug,sanyal}.} allows to
construct a Noether vector. The Noether Symmetry Approach to
reduce dynamics is described in Sec.4. The goal of the method is
to construct a vector field which, contracted with the point-like
Lagrangian, allows to find out, if existing, the conserved
quantities of dynamics. Then it is possible to recast the original
Lagrangian in a new set of variables where cyclic ones explicitly
appear. The number of cyclic variables is equal to the number of
conserved quantities. This technique reduces the order of
derivation of the equations and simplifies the process to achieve
exact solutions. We carry out the method for the point-like $f(R)$
Lagrangian in spherical symmetry  and find out some exact
solutions in Sec.5. Discussion and conclusions are drawn in Sec.6.
Appendix A is devoted to the field equations in spherical symmetry
for a generic $f(R)$. In Appendix B, we write explicitly the PDE
system, derived from the contraction $L_{X}{\cal L}=0$, which is
the existence condition for the Noether Symmetry.

\section{The $f(R)$ gravity action and the field equations}

The action
\begin{equation}\label{ac}
\mathcal{A}=\int d^4x\sqrt{-g}\left[f(R)+{\cal L}_m\right],
\end{equation}
describes a theory of gravity where $f(R)$ is a generic function
of scalar curvature, $g$ is the determinant of the metric tensor
and ${\cal L}_m $ is the standard fluid matter minimally coupled
with gravity. We are assuming physical units $8\pi G=1$. The field
equations in the metric approach (i.e. obtained by a variation
with respect to the metric  $g_{\mu\nu}$) are
\begin{equation}\label{fe}
f_R(R)R_{\mu\nu}-\frac{1}{2}f(R)g_{\mu\nu}-f_R(R)_{;\mu\nu}+g_{\mu\nu}\Box
f_R(R)\,=\,T^m_{\mu\nu}\,,
\end{equation}
where ${\displaystyle f_R(R)=\frac{df(R)}{dR}}$. Eqs.(\ref{fe})
are of fourth order  due to the covariant derivatives of $f_R(R)$
and reduce to the standard Einstein ones if $f(R)=R$.
$T^m_{\mu\nu}$ is the matter fluid stress-energy tensor. Defining
a {\it curvature stress\,-\,energy tensor}
\begin{equation} \label{curva}
T^{curv}_{\mu\nu}\,=\,\frac{1}{f_R(R)}\left\{\frac{1}{2}g_{\mu\nu}\left[f(R)-Rf_R(R)\right]
+f_R(R)^{;\alpha\beta}(g_{\alpha\mu}g_{\beta\nu}-g_{\mu\nu}g_{\alpha\beta})
\right\}\,,
\end{equation}
Eqs.(\ref{fe}) can be recast in the Einstein\,-\,like form\,:
 \begin{equation}\label{5}
G_{\mu \nu} = R_{\mu\nu}-\frac{1}{2}g_{\mu\nu}R =
T^{curv}_{\mu\nu}+T^{m}_{\mu\nu}/f_R(R)
\end{equation}
where  matter non\,-\,minimally couples to geometry through the
term $1/f_R(R)$. In this paper, we shall seek for exact solutions
in vacuum so that  we can assume $T^{m}_{\mu\nu}=0$.

By contracting with respect to the metric tensor, one can reveal
the analogy of $f_R(R)$ with  a scalar field being the trace
equation
\begin{equation}\label{}
3\Box f_R(R)+Rf_R(R)-2f(R)\,=\,0\,.
\end{equation}
It can be seen as  a Klein-Gordon equation for an effective scalar
field if the identifications
$$ \varphi \,\longrightarrow \,f_R(R)\,, \qquad
\frac{dV(\varphi)}{d\phi}\,\longrightarrow\,
\frac{Rf_R(R)-2f(R)}{3}\,,$$ are considered \cite{starobinsky}.

\section{The point-like $f(R)$ Lagrangian in spherical symmetry}

As hinted in the introduction, the aim of this paper is to work
out an approach  to obtain spherically symmetric solutions in
fourth order gravity by means of Noether Symmetries. In order to
develop this approach (the method will be outlined in Sec.4),  we
need to deduce a point-like  Lagrangian from the action
(\ref{ac}). Such a Lagrangian can be obtained by imposing the
spherical symmetry in the field action (\ref{ac}). As a
consequence, the infinite number of degrees of freedom of the
original field theory will be reduced to a finite number. The
technique is based on the choice of a suitable Lagrange multiplier
defined by assuming the Ricci scalar, argument of the function
$f(R)$ in spherical symmetry. Elsewhere, this approach has been
successfully used  for  the FRW metric with the purpose to
find out cosmological solutions \cite{noether-capoz,CQGRug,GRGGae}.\\
In general, a spherically symmetric spacetime  can be described
assuming the metric\,:
\begin{equation}\label{me2}
{ds}^2=A(r){dt}^2-B(r){dr}^2-M(r)d\Omega\,,
\end{equation}
where ${d\Omega}={d\theta}^2+{\sin\theta}^2{d\phi}^2$ is the
angular element. Obviously the conditions $M(r)=r^2$ and
$B(r)=A^{-1}(r)$ are requested to obtain the  standard
Schwarzschild case of GR. This condition is necessary if one wants
to recover the standard measure of a circumference when $r$ is the
radius of a circle. Our goal is to reduce the field action
(\ref{ac}) to a form with a finite degrees of freedom, that is the
canonical action
\begin{equation}\label{}
\mathcal{A}=\int dr\mathcal{L}(A, A', B, B', M, M', R, R')
\end{equation}
where the Ricci scalar $R$ and the potentials   $A$, $B$, $M$ are
the set of independent variables defining the configuration space.
Prime indicates the derivative with respect to the radial
coordinate $r$. In order to achieve the point-like Lagrangian in
this set of coordinates, we write
\begin{equation}\label{lm}
\mathcal{A}=\int
d^4x\sqrt{-g}\biggl[f(R)-\lambda(R-\bar{R})\biggr]\,,
\end{equation}
where $\lambda$ is the Lagrangian multiplier and $\bar{R}$ is the
Ricci scalar expressed in terms of the metric (\ref{me2})
\begin{equation}\label{rs}
\bar{R}=\frac{A''}{AB}+2\frac{M''}{BM}+\frac{A'M'}{ABM}-\frac{A'^2}{2A^2B}-\frac{M'^2}{2BM^2}-\frac{A'B'}{2AB^2}-\frac{B'M'}
{B^2M}-\frac{2}{M}\,,
\end{equation}
which can be recast in the more compact form
\begin{equation}
\bar{R}=R^*+\frac{A''}{AB}+2\frac{M''}{BM}\,,
\end{equation}
where $R^*$ collects first order derivative terms. The Lagrange
multiplier $\lambda$ is obtained by varying  the action (\ref{lm})
with respect to $R$. One gets $\lambda= f_R(R)$. By expressing the
determinant $g$ and $\bar{R}$ in terms of $A$, $B$ and $M$, we
have, from Eq.(\ref{lm}),
\begin{eqnarray}\label{ac1}\mathcal{A}=\int
drA^{1/2}B^{1/2}M\biggl[f-f_{R}\biggl(R-R^*-\frac{A''}{AB}-2\frac{M''}{BM}\biggr)\biggr]=\\
=\int dr
\biggl\{A^{1/2}B^{1/2}M\biggl[f-f_{R}(R-R^*)\biggr]-\biggl(\frac{f_{R}M}{A^{1/2}B^{1/2}}\biggr)'A'-
2\biggl(\frac{A^{1/2}}{B^{1/2}}f_{R}\biggr)
'M'\biggr\}\nonumber\,.\end{eqnarray} The two lines differs for a
divergence term which we discard  integrating by parts. Therefore,
the point-like Lagrangian becomes\,:
\begin{eqnarray}\label{lag}
\mathcal{L}=-\frac{A^{1/2}f_{R}}{2MB^{1/2}}{M'}^2-\frac{f_{R}}{A^{1/2}B^{1/2}}A'M'-\frac{Mf_{RR
}}{A^{1/2}B^{1/2}}A'R'+\nonumber\\\nonumber\\-\frac{2A^{1/2}f_{RR}}{B^{1/2}}R'M'-A^{1/2}B^{1/2}[(2+MR)f_{R}-Mf]\,,
\end{eqnarray}
which is canonical since only the configuration variables and
their first order derivatives with respect to $r$ are present. Eq.
(\ref{lag}) can be recast in more compact form introducing the
matrix formalism\,:
\begin{equation}\label{la}
\mathcal{L}={\underline{q}'}^t\hat{T}\underline{q}'+V
\end{equation}
where $\underline{q}=(A,B,M,R)$ and $\underline{q}'=(A',B',M',R')$
are the generalized positions and velocities associated to
$\mathcal{L}$. The index ``{\it t}" indicates the transposed column
vector. The kinetic tensor is given by ${\displaystyle
\hat{T}_{ij}\,=\,\frac{\partial^2\mathcal{L}}{\partial q'_i\partial
q'_j}}$.  $V=V(q)$ is the potential depending only on the
configuration variables. The Euler\,-\,Lagrange equations read
\begin{eqnarray}\label{fe2}\nonumber
\frac{d}{dr}\nabla_{q'}\mathcal{L}-\nabla_{q}\mathcal{L}=2\frac{d}{dr}\biggl(\hat{T}\underline{q}'\biggr)-\nabla_{q}V-{\underline
{q}'}^t\biggl(\nabla_{q}\hat{T}\biggr)\underline{q}'\,=\nonumber\\\nonumber\\=\,2\hat{T}\underline{q}''+2\biggl(\underline{q}'\cdot\nabla_{q}\hat{T}\biggr)
\underline{q}'-\nabla_{q}V-\underline{q}'^t\biggl(\nabla_{q}\hat{T}\biggr)\underline{q}'=0
\end{eqnarray}
which furnish the  equations of motion in term of  $A$, $B$, $M$
and $R$, respectively. The field equation for $R$ corresponds to
the constraint among the configuration coordinates. It is worth
noting that the Hessian determinant of (\ref{lag}),
${\displaystyle \left|\left|\frac{\partial^2\mathcal{L}}{\partial
q'_i\partial q'_j}\right|\right|}$, is zero. This result clearly
depends on the absence of the generalized velocity $B'$ into the
point\,-\,like Lagrangian. As matter of fact, using a point-like
Lagrangian approach implies that the metric variable $B$ does not
contributes to dynamics, but the  equation
of motion for $B$ has to be considered as a further constraint equation.\\
Beside the Euler\,-\,Lagrange equations (\ref{fe2}),  one has to
take into account the  energy $E_\mathcal{L}$\,:
\begin{equation}\label{ene}
E_\mathcal{L}={\underline
{q}'}\cdot\nabla_{q'}\mathcal{L}-\mathcal{L}
\end{equation}
which can be easily recognized to be coincident with the Euler-Lagrangian equation
for the component $B$ of the generalized position $\underline{q}$.
Then  the Lagrangian (\ref{lag})  has three  degrees of freedom
and not four, as we would expected "a
priori".\\
Now, since the motion equation describing the evolution of the metric potential $B$ does not depends on its derivative,
it can be explicitly solved in term of $B$ as a function of other coordinates\,:
\begin{equation}\label{eqb}
B=\frac{2M^2f_{RR}A'R'+2Mf_{R}A'M'+4AMf_{RR}M'R'+Af_{R}M'^2}{2AM[(2+MR)f_{R}-Mf]}\,.
\end{equation}
By inserting Eq.(\ref{eqb}) into the Lagrangian (\ref{lag}),  we
obtain a non-vanishing Hessian matrix removing the singular
dynamics. The new Lagrangian reads\footnote{Lowering  the dimension
of configuration space through the substitution (\ref{eqb}) does
not affect the  dynamics, since $B$ is a non-evolving quantity. In
fact, introducing Eq. (\ref{eqb}) directly into the dynamical
equations given by (\ref{lag}),  they  coincide with those derived
by (\ref{lag2}).}
\begin{equation}\label{}\mathcal{L}^*= {\bf L}^{1/2}\end{equation}
with
\begin{eqnarray}\label{lag2}\nonumber
{\bf L}=\underline{q'}^t\hat{{\bf
L}}\underline{q'}=\frac{[(2+MR)f_{R}-fM]}{M}\times\nonumber\\\nonumber\\\times[2M^2f_{RR}A'R'
+2MM'(f_{R}A'+2Af_{RR}R') +Af_{R}M'^2]\,.
\end{eqnarray}
Since ${\displaystyle \frac{\partial{\bf L}}{\partial r}=0}$,
${\bf L}$ is canonical (${\bf L}$ is the quadratic form of
generalized velocities, $A'$, $M'$ and $R'$ and then coincides
with the Hamiltonian), so that we can consider ${\bf L}$ as the
new Lagrangian with three  degrees of freedom. The crucial point
of such a replacement is that the Hessian determinant is now
non\,-\,vanishing, being\,:
\begin{equation}\label{}
\left|\left|\frac{\partial^2 {\bf L}}{\partial q'_i\partial
q'_j}\right|\right|=3AM[(2+MR)f_{R}-Mf]^3f_{R}{f_{RR}}^2\,.
\end{equation}
Obviously, we are supposing that $(2+MR)f_{R}-Mf\neq0$, otherwise the
above definitions of $B$, [Eq.(\ref{eqb})], and ${\bf L}$,
[Eq.(\ref{lag2})], lose of significance, besides we assume $f_{RR}\neq 0$
to admit a wide class of fourth order gravity models. The case
$f(R)\,=\,R$  requires a different investigation. In fact,
considering the GR point\,-\,like Lagrangian needs a further
lowering of  degrees of freedom of the system and the previous
results cannot be straightforwardly  considered.
 From (\ref{lag}), we get\,:
\begin{equation}\label{lagr}
\mathcal{L}_{GR}=-\frac{A^{1/2}}{2MB^{1/2}}{M'}^2-\frac{1}{A^{1/2}B^{1/2}}A'M'-2A^{1/2}B^{1/2}\,,
\end{equation}
whose Euler-Lagrange equations  provide the standard equations of
GR for Schwarzschild metric. It is easy to see the absence of the
generalized velocity $B'$ in Eq.(\ref{lagr}). Again, the Hessian
determinant is zero. Nevertheless, considering, as above, the
constraint (\ref{eqb}) for $B$, it is possible to obtain a
Lagrangian  with a non-vanishing Hessian. In particular one has\,:
\begin{equation}\label{eqbgr}B_{GR}=\frac{M'^2}{4M}+\frac{A'M'}{2A}\,,\end{equation}
\begin{equation}\label{lag2gr}\mathcal{L}_{GR}^*= {\bf L}_{GR}^{1/2}=\sqrt{\frac{M'(2MA'+AM')}{M}}\,,\end{equation}
and then the Hessian determinant is
\begin{equation}\label{}\left|\left|\frac{\partial^2 {\bf L}_{GR}}{\partial q'_i\partial q'_j}\right|\right|=-1\,,\end{equation}
which is nothing else but a non-vanishing sub-matrix of the $f(R)$
Hessian matrix.\\
 Considering the Euler\,-\, Lagrange equations
coming from (\ref{eqbgr}) and (\ref{lag2gr}),
 one  obtains the vacuum solutions of  GR, that is\,:
\begin{equation}\label{schsol}
A=k_{4}-\frac{k_{3}}{r+k_{1}}\,,\ \ \ \ \ \ \
B=\frac{k_{2}k_{4}}{A}\,,\ \ \ \ \ \ M=k_{2}(r+k_{1})^2\,.
\end{equation}
In particular, the standard form of Schwarzschild solution is obtained for
$k_{1}=0$, $k_{2}=1$, ${\displaystyle k_{3}=\frac{2GM}{c^2}}$ and
$k_{4}=1$.\\
A formal summary of the field equations descending from the
point\,-\,like Lagrangian and their relation with respect to the
ones of the standard approach is given  in Tab.1.

\begin{table}[h]
\begin{center}
\begin{tabular}{|ccc|} \hline
  Field equations approach &  & Point-like Lagrangian approach \\
  $\downarrow$ &  &  $\downarrow$ \\
  $\delta\int d^4x\sqrt{-g}f=0$  & $\leftrightarrows$ &  $\delta\int dr\mathcal{L}=0$  \\
  $\downarrow$ &  &  $\downarrow$ \\
  $H_{\mu\nu}=\partial_\rho\frac{\partial(\sqrt{-g}f)}{\partial_\rho g^{\mu\nu}}-\frac{\partial (\sqrt{-g}f)}{\partial g^{\mu\nu}}=0$ &
  & $\frac{d}{dr}\nabla_{q'}\mathcal{L}-\nabla_{q}\mathcal{L}=0$ \\  & $\leftrightarrows$ &   \\
  $H=g^{\mu\nu}H_{\mu\nu}=0$ &
  & $E_\mathcal{L}={\underline
  {q}'}\cdot\nabla_{q'}\mathcal{L}-\mathcal{L}$ \\
  $\downarrow$ &  & $\downarrow$  \\
  $H_{00}=0$ & $\leftrightarrows$ & $\frac{d}{dr}\frac{\partial \mathcal{L}}{\partial A'}-\frac{\partial \mathcal{L}}{\partial A}=0$ \\
  $H_{rr}=0$ & $\leftrightarrows$ & $\frac{d}{dr}\frac{\partial \mathcal{L}}{\partial B'}-\frac{\partial \mathcal{L}}{\partial B}\propto E_\mathcal{L}=0$ \\
  $H_{\theta\theta}={\csc^{2}\theta}H_{\phi\phi}=0$ & $\leftrightarrows$ & $\frac{d}{dr}\frac{\partial \mathcal{L}}{\partial M'}-\frac{\partial \mathcal{L}}{\partial M}=0$ \\
  $H=A^{-1}H_{00}-B^{-1}H_{rr}-2M^{-1}{\csc^{2}\theta}H_{\phi\phi}=0$ & $\leftrightarrows$ & A combination of the above equations \\\hline
\end{tabular}
\end{center}
\caption{ The field-equations approach and the point-like
Lagrangian approach differ since the symmetry, in our case the
spherical one, can be imposed whether in the field equations,
after standard variation with respect to the metric, or directly
into the Lagrangian, which becomes point-like. The energy
$E_\mathcal{L}$ corresponds to the $00$\,-\,component of
$H_{\mu\nu}$.  The absence of $B'$ in the Lagrangian implies the
proportionality between the constraint equation for $B$ and the
energy function $E_\mathcal{L}$. As a consequence, the number of
independent equations is three (as the number of unknown
functions). Finally it is obvious the correspondence between
$\theta\theta$ component and field equation for $M$. The explicit
form of field equations $H_{\mu\nu}$ is given in App.B.
}\end{table}

\section{The Noether Symmetry Approach}
In order to find out solutions for the  Lagrangian (\ref{lag2}),
we can search for symmetries related to cyclic variables and then
reduce dynamics. This approach allows, in principle,  to select
$f(R)$ gravity models compatible with spherical symmetry. As a
general remark, the Noether Theorem states that conserved
quantities are related to the existence of cyclic variables into
dynamics \cite{arnold,marmo,morandi}. Let us give a summary of the
approach for finite dimensional dynamical systems.

Let ${\l}(q^{i}, \dot{q}^i)$ be a canonical, non-degenerate
point-like Lagrangian where  \beq \label{01} \frac{\pa {\l}}{\pa
\lambda}=0\,;\;\;\;\;\;\;\; \mbox{det}H_{ij}\eqdef \mbox{det}
\left|\left| \frac{\pa^2 {\l}}{\pa
\dot{q}^i\pa\dot{q}^j}\right|\right|\neq 0\,,  \eeq with $H_{ij}$
as above,  the Hessian matrix related to ${\l}$. The dot indicates
derivatives with respect to the affine parameter $\lambda$ which,
ordinarily, corresponds to time $t$. In our case, it is the radial
coordinate $r$. In standard problems of analytical mechanics,
${\l}$ is in the form \beq \label{02}
{\l}=T({\bq},\dot{\bq})-V({\bq})\;, \eeq where $T$ and $V$ are the
"kinetic"  and "potential energy" respectively. $T$ is a positive
definite quadratic form in $\dot{\bq}$.  The energy function
associated with ${\l}$ is \beq \label{03} E_{\l}\equiv\frac{\pa
{\l}}{\pa \qd^{i}}\qd^i-{\l}\,, \eeq which is  the total energy
$T+V$. It has to be noted that $E_{\l}$ is, in any case, a
constant of  motion. In this formalism, we are going to consider
only transformations which are point-transformations. Any
invertible and smooth transformation of the "positions"
$Q^{i}=Q^{i}({\bq})$ induces a transformation of the "velocities"
such that \beq \label{04} \dot{Q}^i({\bq})=\frac{\pa Q^i}{\pa
q^j}\qd^j\;; \eeq the matrix ${\cal J}=|| \pa Q^i/\pa q^j ||$ is
the Jacobian of the transformation on the positions, and it is
assumed to be nonzero. The Jacobian $\widetilde{{\cal J}}$ of the
"induced" transformation is easily derived and ${\cal J}\neq
0\rightarrow \widetilde{{\cal J}}\neq 0$. Usually, this condition
is not satisfied in the whole space but only in the neighbor  of a
point. It is  {\it local transformation}. If one extends the
transformation to the maximal submanifold  such that ${\cal J}\neq
0$, it is possible to
 get  troubles for the whole manifold
due to  possible different topologies \cite{morandi}.

A point transformation $Q^{i}=Q^{i}(\bq)$ can depend on one (or more than one) parameter. Let us assume that a
point transformation depends on a parameter $\varepsilon$, i.e. $Q^{i}=Q^{i}(\bq,\varepsilon)$, and that it gives
rise to a one--parameter Lie group. For infinitesimal values of $\varepsilon$, the transformation is then
generated by a vector field: for instance, as well known, $\pa/\pa x$ represents a translation along the $x$ axis,
$x(\pa/\pa y)-y(\pa/\pa x)$ is a rotation around the $z$ axis and so on. In general, an infinitesimal point
transformation is represented by a generic vector field on $Q$ \beq \label{04b} {\bf
X}=\alp^i({\bq})\frac{\pa}{\pa q^i}\;. \eeq The induced transformation (\ref{04}) is then represented by \beq
\label{05} {\bf X}^{c}=\alp^{i}({\bq})\frac{\pa}{\pa q^{i}}+
\left(\frac{d}{d\lambda}\alp^{i}({\bq})\right)\frac{\pa}{\pa \qd^j}\;. \eeq ${\bf X}^{c}$ is called the "complete
lift" of ${\bf X}$ \cite{morandi}. A function $f(\bq, \bqd)$ is invariant under the transformation  ${\bf X}^{c}$
if \beq \label{06} L_{{\bf X}^{c}}f\eqdef\alp^{i}({\bq})\frac{\pa f}{\pa q^{i}}+
\left(\frac{d}{d\lambda}\alp^{i}({\bq})\right)\frac{\pa f}{\pa \qd^j}\,=\,0\;, \eeq where $L_{{{\bf X}^{c}}}f$ is
the Lie derivative of $f$. In particular, if $L_{{{\bf X}^{c}}}{\l}=0$, ${\bf X}^{c}$ is said to be a {\it
symmetry} for the dynamics derived by ${\l}$.

In order to see how Noether's theorem and cyclic variables are
related, let us consider a Lagrangian ${\l}$ and its
Euler-Lagrange equations \beq \label{07}
\frac{d}{d\lambda}\frac{\pa {\l}}{\pa\qd^{j}}-\frac{\pa {\l}}{\pa
q^{j}}=0\,. \eeq Let us consider also the vector field
(\ref{05}). Contracting (\ref{07}) with the $\alpha^{i}$'s gives
\beq \label{06a} \alp^{j}\left( \frac{d}{d\lambda}\frac{\pa
{\l}}{\pa \qd^j}- \frac{\pa {\l}}{\pa q^j}\right)=0\,. \eeq Being
\beq \label{06b} \alp^{j}\frac{d}{d\lambda}\frac{\pa {\l}}{\pa
\qd^j}= \frac{d}{d\lambda}\left(\alp^j\frac{\pa {\l}}{\pa \qd
^j}\right)- \left(\frac{d \alp^j}{d\lambda}\right)\frac{\pa
{\l}}{\pa \qd ^j}\,, \eeq from (\ref{06a}), we obtain \beq
\label{08} \frac{d}{d\lambda}\left(\alp^{i}\frac{\pa {\l}}{\pa
\qd^i} \right)=L_{\bf X}{\l}\,. \eeq The immediate consequence is
the {\it Noether Theorem}\footnote{In the following, with abuse of
notation, we shall write ${\bf X}$ instead of ${\bf X}^{c}$,
whenever no confusion is possible.}:

\vspace{3. mm}

\noindent {\it If $L_{\bf X}{\l}=0$, then the function \beq
\label{09} \Sigma_{0}=\alp^{i}\frac{\pa {\l}}{\pa \qd^i} \,, \eeq
is a constant of motion.}

\vspace{3. mm}

\noindent {\bf Remark.}
 Eq.(\ref{09}) can be expressed independently of
 coordinates as a contraction of ${\bf X}$ by a Cartan one--form
\beq \label{09a} \theta_{\l} \eqdef \frac{\pa {\l}}{\pa \qd^i}dq^i
\; . \eeq For a generic vector field $ {\bf Y} = y^i \pa / \pa x^i
$, and one--form $\beta = \beta_i d x^i $, we have, by definition,
$ {\displaystyle i_{\bf Y} \beta = y^i \beta_i} $. Thus
Eq.(\ref{09}) can be written as \beq \label{09b} i_{\bf X} \theta
_{\l} = \Sigma_{0} \; . \eeq By a  point--transformation,   the
vector field ${\bf X}$ becomes \beq \label{09c} \widetilde{{\bf
X}} = (i_{\bf X} d Q^k) \frac{\pa}{\pa Q^k} +
     \left( \frac{d}{d\lambda} (i_x d Q^k)\right) \frac{\pa}{\pa \dot{Q}^k} \; .
\eeq We see that $\widetilde{{\bf X}}'$ is still the lift of a
vector field defined on the "space of positions". If ${\bf X}$ is
a symmetry and we choose a point transformation  such  that \beq
\label{010} i_{\bf X} dQ^1 = 1 \; ; \;\;\; i_{\bf X} dQ^i = 0
\;\;\; i \neq 1 \; , \eeq we get \beq \label{010a} \widetilde{{\bf
X}} = \frac{\pa}{\pa Q^1} \;;\;\;\;\;  \frac{\pa {\l}}{\pa Q^1} =
0 \; . \eeq Thus $Q^1$ is a cyclic coordinate and the dynamics can
be reduced \cite{arnold,marmo}.

\noindent {\bf Remarks\/}:
\begin{enumerate}
\item The change of coordinates defined by (\ref{010}) is not unique. Usually
a clever choice is very important.
\item In general, the solution of Eq.(\ref{010}) is not well defined
on the whole space. It is {\it local\/} in the sense explained
above.
\item It is possible that more than one ${\bf X}$ is found,
say for instance ${\bf X}_1$, ${\bf X}_2$. If they commute, i.e. $
[{\bf X}_1, {\bf X}_2] = 0 $, then it is possible to obtain two
cyclic coordinates by solving the system \beq i_{\bf {X_{1}}} dQ^1
= 1; \,  i_{\bf {X_{2}}} dQ^2 = 1; \, i_{\bf {X_{1}}} dQ^i = 0;\,i
\neq 1;\,  i_{\bf {X_{2}}} dQ^i = 0; \, i \neq 2\,. \eeq The
transformed fields will be $\pa/\pa Q^{1}$, $\pa/\pa Q^{2}$. If
they do not commute, this procedure is clearly not applicable,
since commutation relations are preserved by diffeomorphisms. Let
us note that ${\bf X}_3 = [{\bf X}_1, {\bf X}_2]$ is also a
symmetry, indeed, being
$L_{\bf{X_{3}}}{\l}=L_{\bf{X_{1}}}L_{\bf{X_{2}}}{\l}-
L_{\bf{X_{2}}}L_{\bf{X_{1}}}{\l}=0$. If ${\bf X_{3}}$ is
independent of ${\bf X_{1}}$, ${\bf X_{2}}$, we can go on until
the vector fields close the Lie algebra. The usual way to treat
this situation is to make a Legendre transformation, going to the
Hamiltonian formalism and to a Lie algebra of Poisson brackets. If
we    look for a reduction with cyclic coordinates, this procedure
is possible in the following way:
\begin{itemize}
\item we arbitrarily choose  one of the symmetries,
or a linear combination of them, and get new coordinates as above.
After the reduction, we get a new Lagrangian
$\widetilde{{\l}}({\bf Q})$;
\item we search again for symmetries in this new space, make a new reduction
and so on until possible;
\item if the search fails, we try again with another of the existing
symmetries.
\end{itemize}
\end{enumerate}
Let us now assume that ${\l}$ is of the form (\ref{02}). As ${\bf
X}$ is of the form (\ref{05}), $L_{\bf X}{\l}$ will be a
homogeneous polynomial of second degree in the velocities plus a
inhomogeneous term in the $q^{i}$. Since such a polynomial
 has to be identically zero,
each coefficient must be independently zero. If $n$ is the
dimension of the configuration space, we get $\{1+n(n+1)/2\}$
partial differential equations (PDE). The system is
overdetermined, therefore, if any solution exists, it will be
expressed in terms of integration constants instead of boundary
conditions. It is also obvious that an overall constant factor in
the Lie vector ${\bf X}$ is irrelevant. In other words, the
Noether Symmetry Approach can be used to select
 functions which assign the models and, as we shall see below,
such functions (and then the models) can be physically relevant.
This fact justifies the method at least {\it a posteriori}.

\section{The Noether Approach for $f(R)$ gravity in spherical symmetry }

Since the above considerations, if one assumes the spherical
symmetry, the role of the {\it affine parameter} is  played by the
coordinate radius $r$.  In this case, the configuration space is
given by $\mathcal{Q}=\{A, M, R\}$ and the tangent space by
$\mathcal{TQ}=\{A, A', M, M', R, R'\}$. On the other hand,
according to the Noether theorem, the existence of a symmetry for
dynamics described by the Lagrangian (\ref{lag2})  implies a
constant of motion. Let us apply the Lie derivative to the
(\ref{lag2}), we have\footnote{From now on,  $\underline{q}$
indicates the vector $(A,M,R)$.}\,:
\begin{equation}\label{}
L_{\mathbf{X}}{\bf L}\,=\,\underline{\alpha}\cdot\nabla_{q}{\bf L
}+\underline{\alpha}'\cdot\nabla_{q'}{\bf L}
=\underline{q}'^t\biggl[\underline{\alpha}\cdot\nabla_{q}\hat{{\bf
L }}+ 2\biggl(\nabla_{q}\alpha\biggr)^t\hat{{\bf L
}}\biggr]\underline{q}'\,,
\end{equation}
which vanish if the functions ${\underline{\alpha}}$ satisfy the
following system (see App.B for details)
\begin{equation}\label{sys}
\underline{\alpha}\cdot\nabla_{q}\hat{{\bf L}}
+2(\nabla_{q}{\underline{\alpha}})^t\hat{{\bf L
}}\,=\,0\,\longrightarrow\ \ \ \ \alpha_{i}\frac{\partial
\hat{{\bf L}}_{km}}{\partial
q_{i}}+2\frac{\partial\alpha_{i}}{\partial q_{k}}\hat{{\bf L
}}_{im}=0\,.
\end{equation}
Solving the system (\ref{sys})  means to find out the functions
$\alpha_{i}$ which assign the Noether vector. However the system
(\ref{sys}) implicitly depends on the form of $f(R)$ and then, by
solving it, we get also $f(R)$ theories compatible with spherical
symmetry. On the other hand, by choosing the $f(R)$ form, we can
explicitly solve (\ref{sys}). As an example, one finds that the
system (\ref{sys}) is satisfied if we chose
\begin{equation}\label{solsy}f(R)\,=\,f_0 R^s\ \ \ \ \  \underline{\alpha}=(\alpha_1,\alpha_2,\alpha_3)=
\biggl((3-2s)kA,\ -kM,\ kR\biggr)\,
\end{equation}
with $s$ a real number, $k$ an integration constant and $f_0$ a
dimensional coupling constant\footnote{ The dimensions are given
by $R^{1-s}$ in term of the Ricci scalar. For the sake of
simplicity we will put $f_0=1$ in the forthcoming discussion.}.
This means that for any $f(R)=R^s$ exists, at least, a Noether
symmetry and  a related constant of motion $\Sigma_{0}$\,:
\begin{eqnarray}\label{cm}\nonumber
\Sigma_{0}\,=\,\underline{\alpha}\cdot\nabla_{q'}{\bf L
}=\nonumber\\\nonumber\\=2
skMR^{2s-3}[2s+(s-1)MR][(s-2)RA'-(2s^2-3s+1)AR']\,.\end{eqnarray}
A physical interpretation of $\Sigma_{0}$  is possible if one
gives an interpretation of this quantity in GR. In such a case,
with $s=1$, the above procedure has to be applied to the
Lagrangian (\ref{lag2gr}). We obtain the solution
\begin{equation}\label{solsygr}\underline{\alpha}_{GR}=(-kA,\
kM)\,.
\end{equation}
The functions $A$ and $M$ give the Schwarzschild solution (\ref{schsol}), and then the constant of motion acquires
the form
\begin{equation}\label{cmgr}\Sigma_{0}= \frac{2GM}{c^2}\,.\end{equation}
In other words, in the case of Einstein gravity, the Noether
symmetry gives as a conserved quantity   the Schwarzschild radius
or  the mass of the gravitating system.

Another solution can be find out for $R=R_0$ where $R_0$ is a
constant (see also \cite{multamaki}).  In this case, the field
equations (\ref{fe}) reduce to \beq \label{fe1} R_{\mu\nu}+k_0
g_{\mu\nu}=0\,,\eeq where ${\displaystyle
k_0=-\frac{1}{2}f(R_0)/f_R(R_0)}$. The general solution is
\beq\label{scwdes}
A(r)=\frac{1}{B(r)}=1+\frac{k_0}{r}+\frac{R_0}{12}r^2\,,\qquad
M=r^2\eeq with the special case \beq
A(r)=\frac{1}{B(r)}=1+\frac{k_0}{r}\,,\qquad M=r^2\,,\qquad
R=0\,.\eeq The solution (\ref{scwdes}) is the well known
Schwarzschild-de Sitter one which is a solution in most of
modified gravity theories. It evades the Solar System constraints
due to the smallness of the effective cosmological constant.
However, other spherically symmetric solutions, different from
this, are more significant for Solar System tests.

%+\frac{k_1}{r^2}

In the general case $f(R)=R^s$, the Lagrangian (\ref{lag2})
becomes
\begin{eqnarray}\label{}\nonumber {\bf L}=\frac{sR^{2s-3}[2s+(s-1)MR]}{M}\times\nonumber\\
\nonumber\\\times[2(s-1)M^2A'R'+2MRM'A'+4(s-1)AMM'R'+ARM'^2]\,,\end{eqnarray}
and the expression (\ref{eqb}) for $B$ is
\begin{equation}\label{}B=\frac{s[2(s-1)M^2A'R'+2MRM'A'+4(s-1)AMM'R'+ARM'^2]}{2AMR[2s+(s-1)MR]}\end{equation}
As it can be easily checked, GR is recovered when $s=1$.

Using the constant of motion (\ref{cm}),  we solve in term of $A$
and obtain
\begin{equation}\label{}A=R^{\frac{2s^2-3s+1}{s-2}}\biggl\{k_1+\Sigma_{0}\int\frac{R^{\frac{4s^2-9s+5}{2-s}}dr}{2ks(s-2)M[2s+(s-1)MR]}\biggr\}\end{equation}
for $s\neq2$, with $k_1$ an integration constant. For $s\,=\,2$,
one finds
\begin{equation}\label{}A=-\frac{\Sigma_{0}}{12kr^2(4+r^2R)RR'}\,.\end{equation}
These relations allow to find out general solutions for the field
equations giving the dependence of the Ricci scalar on the radial
coordinate $r$. For example, a solution is found for
\begin{equation}\label{}
s=5/4\,,\ \ \ \ M=r^2\,,\ \ \ \ R= 5 r^{-2}\,,
\end{equation}
obtaining  the spherically symmetric metric
\begin{equation}\label{}ds^2=\frac{1}{\sqrt{5}}(k_2+k_1 r)dt^2-
\frac{1}{2}\biggl(\frac{1}{1+\frac{k_2}{k_1r}}\biggr)dr^2-r^2d\Omega\,,\end{equation}
with ${\displaystyle k_2=\frac{32\Sigma_0}{225 k}}$. It is worth
noting that such exact solution is in the range of $s$ values
ruled out by Solar System observations, as pointed out in
\cite{baclif}.

\section{Discussion and Conclusions}
In this paper, we have discussed a general method to find out
exact solutions in Extended Theories of Gravity when a spherically
symmetric background is taken into account. In particular, we have
searched for exact spherically symmetric solutions in $f(R)$
gravity by asking for the existence of Noether symmetries. We have
developed a general formalism and  given some examples of exact
solutions. The procedure consists in: $i)$ considering the
point-like $f(R)$ Lagrangian where spherical symmetry has been
imposed; $ii)$ deriving the Euler-Lagrange equations; $iii)$
searching for a Noether vector field; $iv)$ reducing dynamics and
then integrating the equations of motion using conserved
quantities. Viceversa, the approach allows also to select families
of $f(R)$ models where a particular symmetry (in this case the
spherical one) is present. As examples, we discussed power law
models and models with constant Ricci  curvature scalar. However,
the above method can be further generalized. If a symmetry exists,
the Noether Approach allows, as discussed in Sec.4,
transformations of variables where the cyclic ones are evident.
This fact allows to reduce dynamics and then  to get more easily
exact solutions. For example, since we know that
$f(R)=R^s$\,-\,gravity admit a conserved quantity, a coordinate
transformation can be induced by the Noether symmetry. We  ask for
the coordinate transformation\,:
\begin{equation}
{\bf L}={\bf L}(\underline{q}, \underline{q}')={\bf L}(A, M, R,
A', M', R')\rightarrow\widetilde{{\bf L }}=\widetilde{{\bf L
}}(\widetilde{M}, \widetilde{R}, \widetilde{A}', \widetilde{M}',
\bar{R}')\,, \end{equation} for the Lagrangian (\ref{lag2}), where
the Noether symmetry, and then the conserved quantity, corresponds
to the cyclic variable $\widetilde{A}$. If more than one symmetry
exists, one can find more than one cyclic variables. In our case,
if three Noether symmetries exist, we can transform the Lagrangian
${\bf L}$ in a Lagrangian with three cyclic coordinates, that is
$\widetilde{A}=\widetilde{A}({\underline{q}})$,
$\widetilde{M}=\widetilde{M}({\underline{q}})$ and
$\widetilde{R}=\widetilde{R}({\underline{q}})$ which are function
of the old ones. These new functions have to  satisfy the
following system
\begin{equation}\label{sys2}
\left\{\begin{array}{ll}
(3-2s)A\frac{\partial\widetilde{A}}{\partial
A}-M\frac{\partial\widetilde{A}}{\partial
M}+R\frac{\partial\widetilde{A}}{\partial R}=1\,,
\\\\
(3-2s)A\frac{\partial\widetilde{q}_i}{\partial
A}-M\frac{\partial\widetilde{q}_i}{\partial
M}+R\frac{\partial\widetilde{q}_i}{\partial R}\,=\,0\,,
\end{array} \right.
\end{equation}
with $i=2,3$ (we have put $k=1$). A solution of (\ref{sys2}) is
given by the set (for $s\neq 3/2$)
\begin{equation}\label{so1}
\left\{ \begin{array}{ll} \widetilde{A}=\frac{\ln
A}{(3-2s)}+F_{A}(A^{\frac{\eta_A}{3-2s}}M^{\eta_A},A^{\frac{\xi_A}{2s-3}}M^{\xi_A})
\\\\
\widetilde{q}_i=F_{i}(A^{\frac{\eta_i}{3-2s}}M^{\eta_i},A^{\frac{\xi_i}{2s-3}}M^{\xi_i})
\end{array} \right.
\end{equation}
and if $s=3/2$
\begin{equation}\label{so2}
\left\{ \begin{array}{ll} \widetilde{A}=-\ln M+F_{A}(A)G_A(MR)
\\\\
\widetilde{q}_i=F_{i}(A)G_i(MR)
\end{array} \right.
\end{equation}
where $F_A$, $F_i$, $G_A$ and $G_i$ are arbitrary functions  and
$\eta_A$, $\eta_i$, $\xi_A$ and $\xi_i$ integration constants.

These considerations show that the Noether Symmetries Approach can
be applied to large classes of gravity theories.  Up to now the
Noether symmetries Approach has been  worked out in the case of
FRW\,-\,metric. In this paper, we have concentrated our attention
to the development of the general formalism in the case of
spherically symmetric spacetimes. Therefore the fact that, even in
the case of a spherical symmetry, it is possible to achieve exact
solutions seems to suggest that this technique can represent a
paradigmatic approach to work out exact solutions in any theory of
gravity. At this stage, the systematic search for exact solution
is well beyond the aim of this paper. A more comprehensive
analysis in this sense  will be the argument of forthcoming
studies. A final comment deserves the possible relevance of this
approach for the above mentioned Birkhoff-Jensen theorem. The
validity of such a theorem is crucial in every theory of gravity,
due to the fact that it is directly related to the physical
properties of self-gravitating systems (stability, stationarity,
etc.). The  results presented in this paper point out that it does
not hold in general for the specific $f(R)$ theories considered.
However, the above technique could be a good approach to select
suitable classes of theories where such a theorem holds.

\appendix

\section{The $f(R)$ field equations in spherical symmetry}

The field equations (Tab.1)  in  spherical symmetry, derived from
the variational principle of the action (\ref{ac}), are

\begin{eqnarray}\label{}H_{00}=2A^2B^2Mf+\{BMA'^2-A[2BA'M'+M(2BA''-A'B')]\}f_{R}
+\nonumber\\\nonumber\\+(-2A^2MB'R'+4A^2BM'R'+4A^2BMR'')f_{RR}+
\nonumber\\\nonumber\\+4A^2BMR'^2f_{RRR}=0\,,\end{eqnarray}

\begin{eqnarray}\label{}H_{rr}=2A^2B^2M^2f+(BM^2A'^2+AM^2A'B'+2A^2MB'M'+2A^2BM'^2+\nonumber\\
\nonumber\\-2ABM^2A''-4A^2BMM'') f_{R}+(2ABM^2A'R'+
\nonumber\\\nonumber\\+4A^2BMM'R')f_{RR}=0\,,\end{eqnarray}

\begin{eqnarray}\label{}H_{\theta\theta}=2AB^2Mf+(4AB^2-BA'M'+AB'M'-2ABM'')f_{R}+\nonumber\\\nonumber\\+(2BMA'R'-2AMB'R'+2ABM'R'+4ABMR'')
f_{RR}+ \nonumber\\\nonumber\\+4ABMR'^2f_{RRR}=0\,,\end{eqnarray}

\begin{eqnarray}H_{\phi\phi}=\sin^2\theta H_{\theta\theta}=0\,.\end{eqnarray}

The trace equation is

\begin{eqnarray}\label{}H=g^{\mu\nu}H_{\mu\nu}=4AB^2Mf-2AB^2MRf_{R}+3(BMA'R'-AMB'R'+\nonumber\\\nonumber\\+2ABM'R'
+2ABMR'')f_{RR}+6ABMR'^2f_{RRR}=0
\end{eqnarray}

\section{The  Noether vector}

The system (\ref{sys}) comes out from  the condition
$L_{\mathbf{X}}{\bf L}=0$ for the existence of the Noether
symmetry.  Considering the configuration space
$\underline{q}=(A\,,M\,,R\,)$ and defining the Noether vector
components $\underline{\alpha}=(\alpha_1\,,\alpha_2\,,\alpha_3)$,
the system (\ref{sys}) assumes the explicit form

\begin{eqnarray}\xi\biggl(\frac{\partial\alpha_2}{\partial A}f_{R}+M\frac{\partial\alpha_3}{\partial A}f_{RR}\biggr)=0
\end{eqnarray}

\begin{eqnarray}\label{}\frac{A}{M}\biggl[(2+MR)\alpha_3 f_{RR}-\frac{2\alpha_2}{M}f_{R}\biggr]f_{R}+\nonumber\\\nonumber\\
+\xi\biggl[\biggl(\frac{\alpha_1}{M}+2\frac{\partial\alpha_1}
{\partial M}+\frac{2A}{M}\frac{\partial\alpha_2}{\partial
M}\biggr)f_{R}+A\biggl(\frac{\alpha_3}{M}+4\frac{\partial \alpha_3
}{\partial M}\biggr)f_{RR}\biggr]=0
\end{eqnarray}

\begin{eqnarray}\xi\biggl(M\frac{\partial \alpha_1}{\partial R}+2A\frac{\partial \alpha_2}{\partial R}\biggr)f_{RR}=0\end{eqnarray}

\begin{eqnarray}\label{}\alpha_2(f-Rf_{R})f_{R} -\xi\biggl[\biggl(\alpha_3+M\frac{\partial\alpha_3}
{\partial M}+2A\frac{\partial\alpha_3}{\partial
A}\biggr)f_{RR}+\nonumber\\\nonumber\\
+\biggl(\frac{\partial\alpha_2}{\partial
M}+\frac{\partial\alpha_1}{\partial A}+\frac{A}{M}\frac{\partial
\alpha_2 }{\partial A}\biggr)f_{R}\biggr\}=0
\end{eqnarray}

\begin{eqnarray}[M(2+MR)\alpha_3 f_{RR}-2\alpha_2 f_{R}]f_{RR}+\xi\biggl[f_{R}\frac{\partial \alpha_2}{\partial
R}+\nonumber\\\nonumber\\
+\biggl(2\alpha_2+M\frac{\partial\alpha_1}{\partial A}+2
A\frac{\partial\alpha_2}{\partial A}+M\frac{\partial
\alpha_3}{\partial R}\biggr)\alpha_3 f_{RR}+Mf_{RRR}\biggr\}=0
\end{eqnarray}

\begin{eqnarray}2A[(2+MR)\alpha_3 f_{RR}-(f-Rf_{R})\alpha_2]f_{RR}+\xi\biggl[\biggl(\frac{\partial\alpha_1}{\partial
R}+\frac{A}{M}\frac{\partial\alpha_2}{\partial
R}\biggr)f_{R}+\nonumber\\\nonumber\\
+\biggl(2\alpha_1 +2A\frac{\partial\alpha_3}{\partial
R}+M\frac{\partial \alpha_1} {\partial M}+2A\frac{\partial
\alpha_2}{\partial M}\biggr)f_{RR}+2A\alpha_3 f_{RRR}\biggr]=0
\end{eqnarray}

with the condition $\xi=(2+MR)f_{R}-Mf\neq 0$, otherwise the
Hessian of Lagrangian ${\bf L}$ (\ref{lag2}) is vanishing.


\section*{References}



\begin{thebibliography}{99}

\bibitem{starobinsky}
Starobinsky A.A. 1980, Phys. Lett. B, 91, 99

\bibitem{kerner}
Kerner R. 1982, Gen. Rel. Grav., 14, 453

\bibitem{noi}
Capozziello S. 2002, Int. J. Mod. Phys. D, 11, 483.\\
Capozziello S., Carloni S., Troisi A. 2003, Rec. Res. Dev. in
Astron. and Astroph., 1, 1, (arXiv\,:\,astro\,-\,ph/0303041).\\
Odintsov S.D., Nojiri S. 2003, Phys. Lett. B, 576, 5\\
Capozziello S., Cardone V.F., Carloni S., Troisi A. 2003, Int. J.
Mod. Phys. D, 12, 1969.\\
Carroll S.M., Duvvuri V., Trodden M., Turner M. 2004, Phys. Rev.
D, 70, 043528.\\ Allemandi G., Borowiec A., Francaviglia M. 2004,
Phys. Rev. D, 70, 103503.\\Nojiri S. and Odintsov S.D. 2004, Gen.
Rel. Grav. 36, 1765.\\ Cognola G., Elizalde E., Nojiri S., S.D.
Odintsov, Zerbini S. 2005, JCAP, 010.

\bibitem{mnras}
Capozziello S., Cardone V.F., Carloni S., Troisi A. 2004, Phys.
Lett. A, 326, 292.\\
Capozziello S., Cardone V.F.,  Troisi A., 2007  \mnras {\bf 375},
1423.


\bibitem{jcap}
Capozziello S., Cardone V.F., Troisi A., 2006,
  JCAP {\bf 08}, 001.


\bibitem{navarro}
Navarro I., Van Acoleyen I., arXiv:astro-ph/0611127.

\bibitem{bean}
Bean R., {\it et al.}, arXiv:astro-ph/0611321.

\bibitem{barrow}
Li B., Barrow J., arXiv:astro-ph/0701111.

\bibitem{mimicking}
Capozziello S., Cardone V.F., Troisi A. 2005, Phys. Rev. D, 71,
043503

\bibitem{noi-odin}
Capozziello S., Nojiri S., Odintsov S.D., Troisi A., 2006,  Phys.
Lett. B 639, 135.

\bibitem{amendola}
Amendola L., Polarski D., Tsujikawa S., 2006,
arXiv:astro-ph/0603703,\\
Amendola L.,  Gannouji R., Polarski D., Tsujikawa S., 2006,
arXiv:astro-ph/0612180.

\bibitem{arturo}
Capozziello S., Troisi A. 2005, Phys. Rev. D, 72, 044022

\bibitem{matteo}
Allemandi G., Francaviglia M., Ruggiero M.L., Tartaglia A., 2005,
Gen. Rel. Grav. 37, 1891;\\
 Ruggiero M.L., Iorio L., 2007, JCAP 0701, 010.


\bibitem{tegmark}
Faulkner T., Tegmark M., Bunn E.F., Mao Y.,  arXiv:astro-ph/0612569 .

\bibitem{olmo}
Olmo G., 2005, Phys.Rev.D 72, 083505.

\bibitem{kamionkowski}
Erickcek A.L., Smith T. L., Kamionkowski M., 2006, Phys.Rev. D74, 121501.

\bibitem{faraoni}
Faraoni V., Nadeau S., 2007, Phys. Rev. D 75, 023501.

\bibitem{stelle}
Stelle K. 1978, Gen. Rel. Grav., 9, 353.

\bibitem{multamaki}
Multam\"aki T., Vilja I. \pr {\bf D 74}, 064022 (2006).

\bibitem{multamaki2}
Multam\"aki T., Vilja I.  (2006) arXiv:astro-ph/0612775 (2006).

\bibitem{deritis}
de Ritis R., Marmo G., Platania G., Rubano C., Scudellaro P.,
Stornaiolo C., 1990, \pr {\bf 42D}, 1091.

\bibitem{noether-capoz}
 Capozziello S. and  de Ritis R., 1993, \pl {\bf A 177}, 1.\\
Capozziello S., de Ritis R., Rubano C., Scudellaro P., 1996, La
Riv. del Nuovo Cimento, {\bf 19},  4, 1.

\bibitem{ellis}
Hawking S.W. and Ellis G.F.R., 1973 {\it The large scale structure
of space-yime}, Cambridge Univ. Press, Cambridge, UK.


\bibitem{clifton}
Clifton T., PhD\,-\,thesis, arXiv:gr-qc/0610071.



\bibitem{neville}
Neville D., 1980, Phys. Rev. {\bf D 21}, 10.

\bibitem{yasskin}
 Ramaswamy S. and  Yasskin P.B. 1979,  Phys. Rev. {\bf D 19},
 2264.

 \bibitem{barraco}
D. E. Barraco and V. H. Hamity, Phys. Rev., 1998, {\bf D 57}, 954;\\
 D. E. Barraco D.E. and  Hamity V.H., 2000,  Phys. Rev. {\bf D
62}, 044027.

\bibitem{baclif}
Clifton T. and Barrow J.D., 2005, Phys. Rev. D {\bf 72}, 103005,\\
Clifton T. and Barrow J.D., 2006, \cqg {\bf 23},2951;\\
Clifton T. and Barrow J.D., 2006, \cqg {\bf 23}, L1.

\bibitem{CQGRug}
Capozziello S. and  de Ritis R., 1994, \cqg {\bf 11}, 107.

\bibitem{sanyal}
Sanyal A.K., Rubano C., Piedipalumbo E., 2003, \grg {\bf 35},
1617.\\
Kamilya S., Modak B., and Biswas S., 2004, \grg {\bf 36}, 661.

\bibitem{GRGGae}
Capozziello S. and  Lambiase G., 2000,  \grg {\bf 32}, 295.




\bibitem{arnold} V.I. Arnold, {\it Mathematical Methods of
Classical Mechanics}, Springer--Verlag, Berlin (1978).

\bibitem{marmo} G. Marmo, E.J. Saletan, A. Simoni and B. Vitale,
{\it Dynamical Systems. A Differential Geometric Approach to
Symmetry and Reduction}, Wiley, New York (1985).

\bibitem{morandi}
Morandi G., Ferrario C., Lo Vecchio G., Marmo G., Rubano C., 1990,
Phys. Rep. 188, 149


\end{thebibliography}
\end{document}